\documentclass[twocolumn,amsmath,amssymb,tighten,prl]{revtex4-1}

\usepackage{graphicx}
\usepackage{dcolumn} 
\usepackage{bm, bbm}      
\usepackage{curves}
\usepackage{epic}
\usepackage{wasysym}
\usepackage[dvips]{color}
\usepackage{epsf, times}
\usepackage{subfigure}
\usepackage{amsthm}

\def\openone{\leavevmode\hbox{\small1\kern-4.2pt\normalsize1}}
\def\DTO{Dy$_2$Ti$_2$O$_7$}
\def\HTO{Ho$_2$Ti$_2$O$_7$}

\def\rnn{r_{\rm nn}}

\def\etal{\textit{et al}.}

\newcommand{\beq}{\begin{equation}}
\newcommand{\eeq}{\end{equation}}
\newcommand{\bea}{\begin{eqnarray}}
\newcommand{\eea}{\end{eqnarray}}
\newcommand{\bfig}{\begin{figure}}
\newcommand{\efig}{\end{figure}}

\begin{document}

\title{
Magnetic Coulomb Fields of Monopoles in Spin Ice and Their Signatures 
in the Internal Field Distribution
      }

\author{
G. Sala$^{1}$
}
\author{
C. Castelnovo$^{1}$
}
\author{
R. Moessner$^{2,*}$
       }
\author{
S. L. Sondhi$^3$
       }
\author{
K. Kitagawa$^4$
       }
\author{
M. Takigawa$^4$
       }
\author{
R. Higashinaka$^{5,6}$
       }
\author{
Y. Maeno$^5$
       }
\affiliation{
$^1$
SEPnet and Hubbard Theory Consortium, Department of Physics,
Royal Holloway University of London,
Egham TW20 0EX, United Kingdom}
\affiliation{
$^2$
Max-Planck-Institut f\"ur Physik komplexer Systeme,
Dresden, 01187, Germany
            }
\affiliation{
$^3$
Department of Physics, Princeton University,
Princeton, NJ 08544, USA
            }
\affiliation{
$^4$
Institute for Solid State Physics, University of Tokyo
5-1-5 Kashiwanoha, Kashiwa, Chiba 277-8581, Japan
            }
\affiliation{
$^5$
Department of Physics, Graduate School of Science, Kyoto University,
Kyoto 606-8502, Japan
            }
\affiliation{
$^6$
present address: Graduate School of Science, Tokyo Metropolitan University,
Hachioji, Tokyo 192-0397, Japan
            }

\date{\today}

\begin{abstract}
Fractionalisation -- the breaking up of an apparently indivisible microscopic 
degree of freedom -- is one of the most counterintuitive phenomena in 
many-body physics. Here we study its most fundamental manifestation in 
spin ice, the only known fractionalised magnetic compound in 3D: we directly 
visualise the $1/r^2$ magnetic Coulomb field of monopoles which emerge as the 
atomic magnetic dipoles fractionalise. We analyse the internal magnetic field 
distribution, relevant for local experimental probes. In particular, 
we present new zero-field NMR measurements which exhibit excellent agreement 
with the calculated lineshapes, noting that this experimental technique can in 
principle measure directly the monopole density in spin ice. The distribution 
of field strengths is captured by a simple analytical form which exhibits a 
low density of low-field sites---in apparent disagreement with reported 
$\mu$SR results. Counterintuitively, the density of low-field locations 
decreases as the local ferromagnetic correlations imposed by the ice rules 
weaken.
\end{abstract}

\maketitle
%
%

\textit{Introduction} --- 
The magnetic field set up by a spin configuration is the most
direct manifestation of the underlying magnetic moments. The discovery
of a new spin state thus holds the promise of generating -- and revealing its
existence in -- novel properties of the field it sets up. 

A case in point is  
spin ice~\cite{Bramwell2001}, which -- uniquely among magnetic materials in 
three dimensions -- exhibits an emergent gauge field and magnetic monopole
excitations~\cite{Castelnovo2008}, which have analogies 
in magnetic  nanoarrays~\cite{ASIgeneral1,ASIgeneral2,ASIgeneral6,Mengotti2011}.
The spin ice state has the great advantage of exhibiting phenomena of 
fundamental conceptual importance  in a setting which as we describe below is 
simple enough to be easily and intuitively visualised:
an order of topological nature manifests itself in the   fractionalisation of 
the microscopic {\em di}pole degrees of freedom, leading to the deconfined 
magnetic {\em mono}poles~\cite{Castelnovo2012}. 

Neutron scattering experiments, which provide magnetic field correlations in
 reciprocal space, have produced some of the strongest evidence so far for
the gauge structure~\cite{Fennell2009,Kadowaki2009} and 
`Dirac strings'~\cite{Morris2009} that emerge at low temperatures.
Another probe which has prominently been employed is muon spin
rotation~\cite{Lago2007,Bramwell2009,Dunsiger2011} ($\mu$SR), which like NMR
is sensitive to the local fields in real space.
For such local probes, studies of the level of detail characteristic of the
neutron analysis are still lacking.

We remedy this situation by computing the spatially resolved
distribution of internal fields in spin ice. Most fundamentally, according to
Ref.~\onlinecite{Castelnovo2008}, the internal
fields in spin ice contain a contribution from the underlying magnetic
monopoles. Isolating and identifying this contribution is thus of great
conceptual importance in corroborating the peculiar nature of these unique
elementary excitations. 

Here we show how to visualise the monopole
contribution: by measuring the field strength at the
considerably-sized magnetic voids of the lattice we find a radially symmetric
signal (Fig.~\ref{fig: super tetrahedra}) which is well-described by the
Coulomb law, $\propto 1/r^2$, with a coefficient that is in good agreement
with the theoretical prediction~\cite{Castelnovo2008}. Even if
measuring the field strength deep inside the sample may be beyond the reach of
current experiments, the Coulomb field of a magnetic monopole near the sample
surface could be accessible to a sufficiently spatially resolved measurement.

\begin{figure}[]
\includegraphics[width=0.4\columnwidth]
                {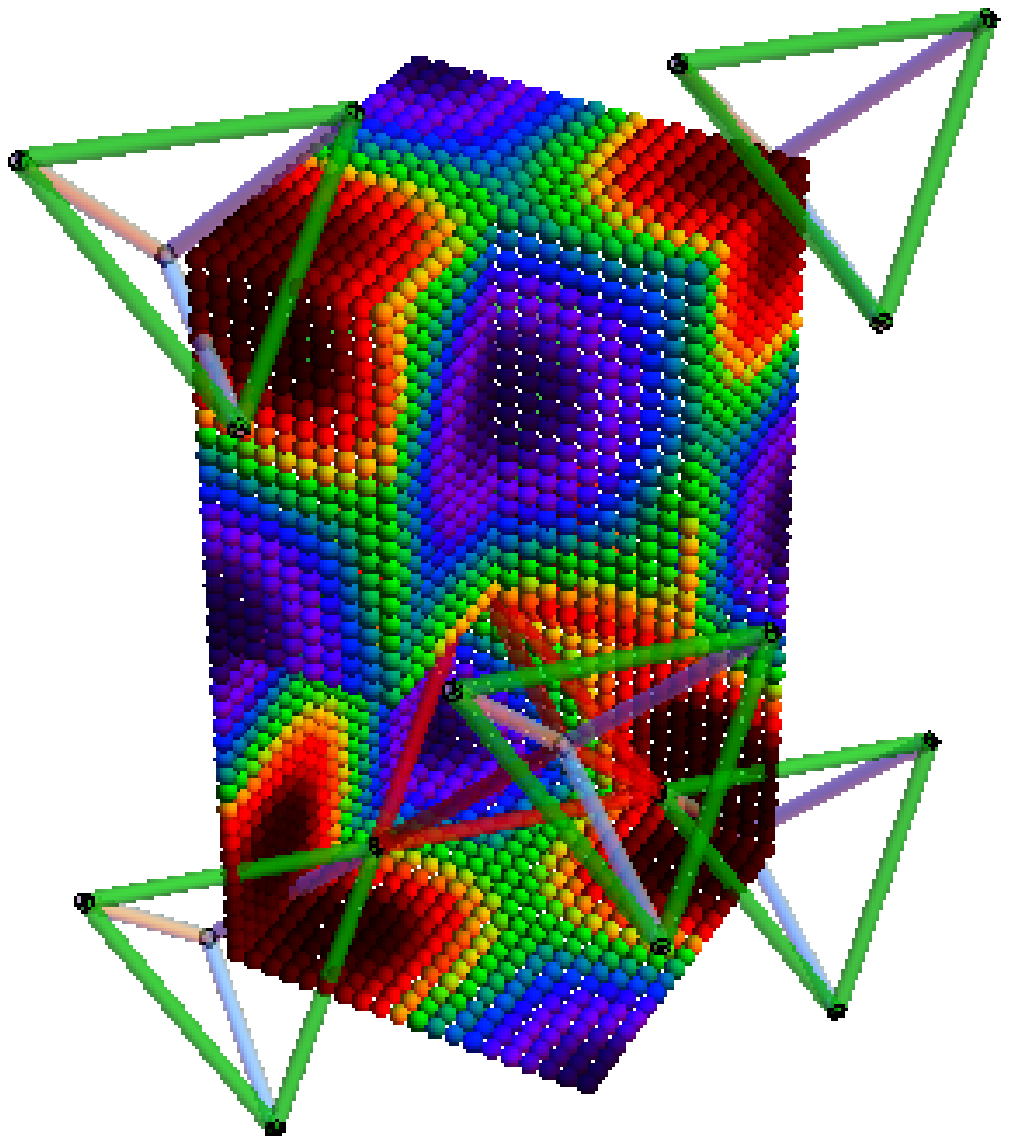}
\includegraphics[width=0.4\columnwidth]
                {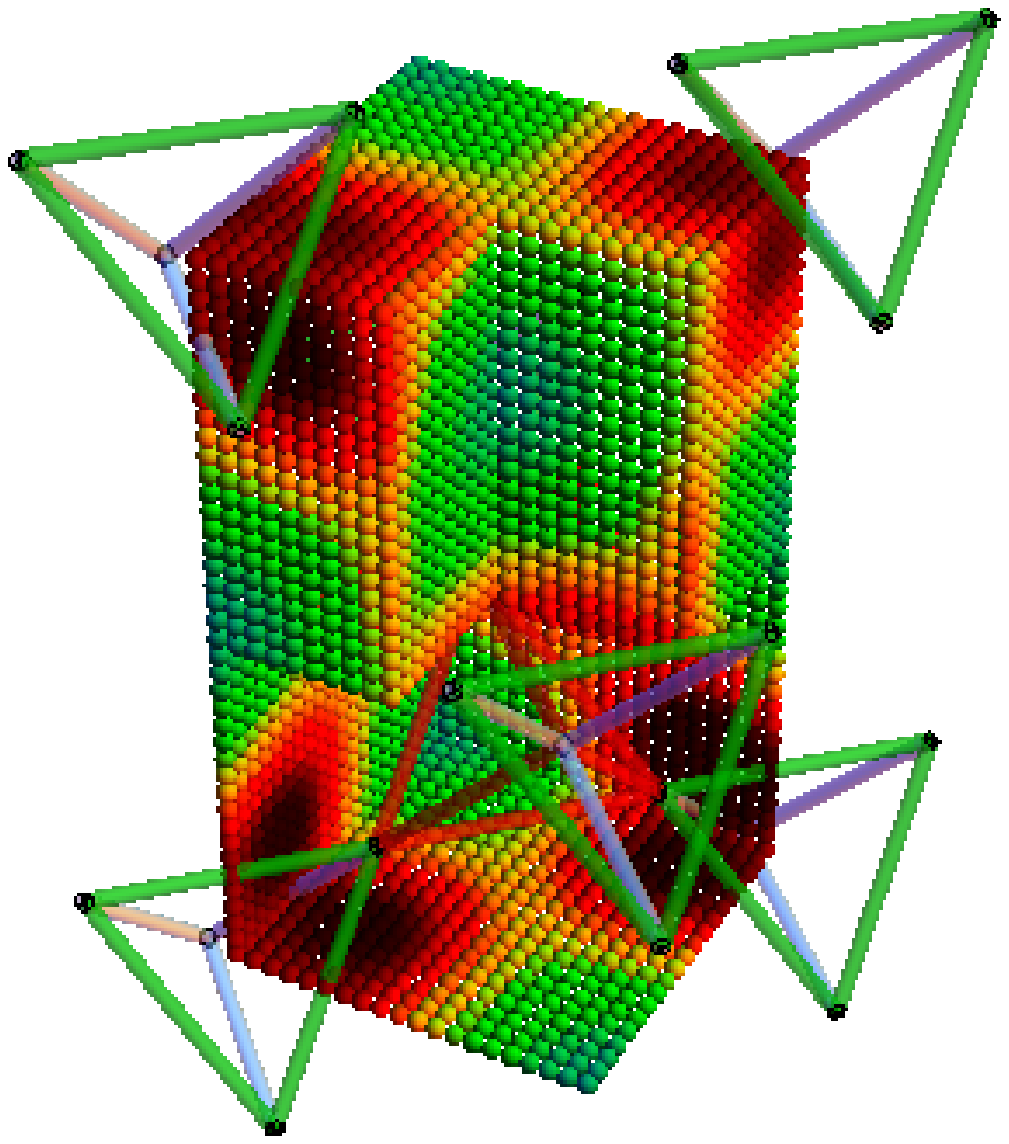}
\includegraphics[width=0.12\columnwidth]
                {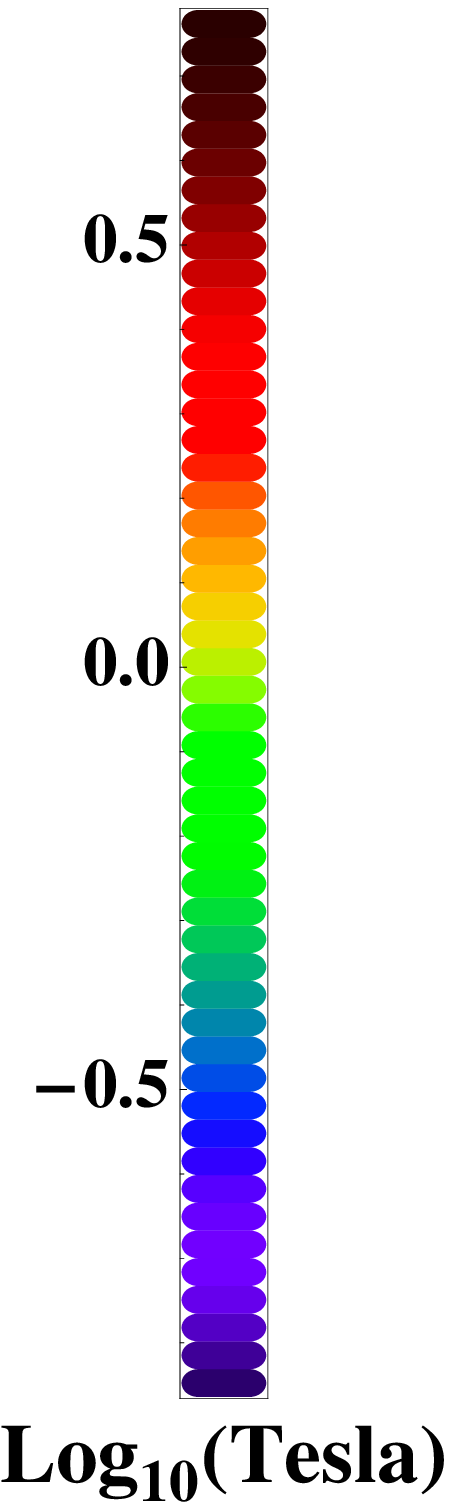}
\caption{
\label{fig: cross section}
Cross sections of the primitive unit cell. 
Left: Average field strength in 2in-2out spin ice configurations. 
The logarithmic colour scale ranges from dark blue ($0.137$~Tesla, the 
smallest average field strength recorded) to deep red ($6$~Tesla). 
Right: Average field strength in completely disordered Ising spin 
configurations.
}
\end{figure}

\begin{figure}[!ht]
\includegraphics[width=0.95\columnwidth,height=0.45\columnwidth]
                {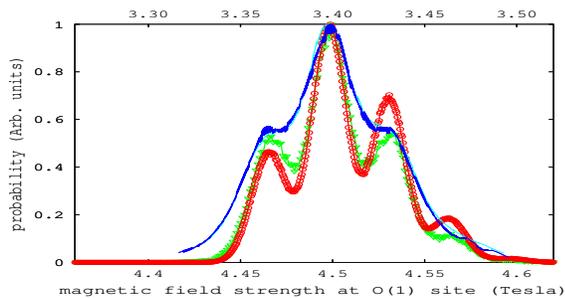}
\caption{
\label{fig: O(1) fields}
Histograms of the magnetic field strength at the O(1) sites obtained from: 
experimental NMR data nominally at $T=0.1$~K (blue) and at $T=0.4$~K (cyan) 
-- top axis; Monte Carlo simulations in equilibrium at $T=0.6$~K (green) and 
equally weighted ensemble of 2in-2out spin ice configurations 
(red) -- bottom axis. 
The experimental curves have been shifted so as to match the main peak 
position from numerics (see the Methods Section in the Supplemental 
Information); the vertical axis is chosen to set the maxima equal to unity. 
}
\end{figure}

To make contact with $\mu$SR and NMR experiments, we compute the full
field distribution in the unit cell
(Fig.~\ref{fig: cross section}). This provides detailed predictions
for NMR experiments, with the lineshape (Fig.~\ref{fig: O(1) fields}) in 
excellent agreement with the first zero-field NMR measurements, the results of 
which we report here. 

Our analysis  places strong constraints on the $\mu$SR signatures
of the spin ice state. In particular, it seems highly unlikely that the signal
detected in Ref.~\onlinecite{Bramwell2009} is due to muons implanted in
pristine bulk spin ice. 
(Our results are consistent with earlier estimates of the internal field 
strength at the muon site~\cite{Dunsiger1996,Lago2007,Dunsiger2011}.)

\begin{figure}[!ht]
\includegraphics[width=0.98\columnwidth,height=0.55\columnwidth]
                {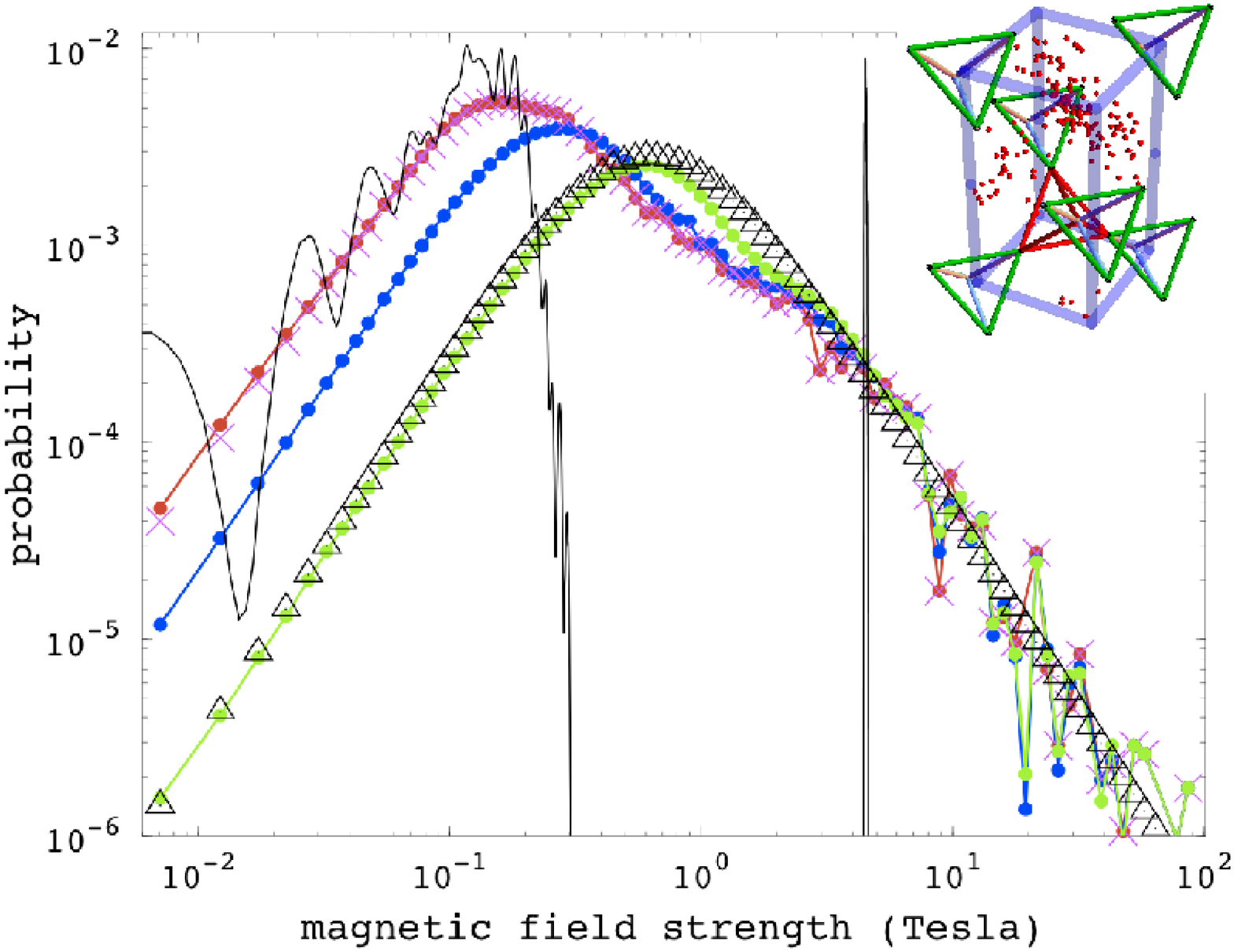}
\caption{
\label{fig: histograms}
Histograms of the field strengths across a uniform cubic grid spanning the
primitive unit cell in a system of size $L=4$ containing $16L^3=1024$ spins.
Red: without monopoles.
Magenta and blue: with two monopoles (see text).
Green: random spin ice configuration (i.e., $T$ much larger than any
interaction energy scale).
Black lines: centres of the super tetrahedra and of the rare earth tetrahedra
(low and high field curves, respectively).
Black triangles: $P(h) \sim h^2 / ( h^2 + H^2_0 )^2$ fit to the random spin
ice behaviour. 
Inset: Spatial distribution of locations of field strength smaller than 
$10$~mTesla in \emph{at least} one of the 10,000 statistically independent 
configurations sampled. 
The dimensionless volume fraction of such sites is approximately 
$5\,10^{-5}$. 
}
\end{figure}

\begin{figure}[ht]
\includegraphics[width=0.95\columnwidth,height=0.7\columnwidth]
                {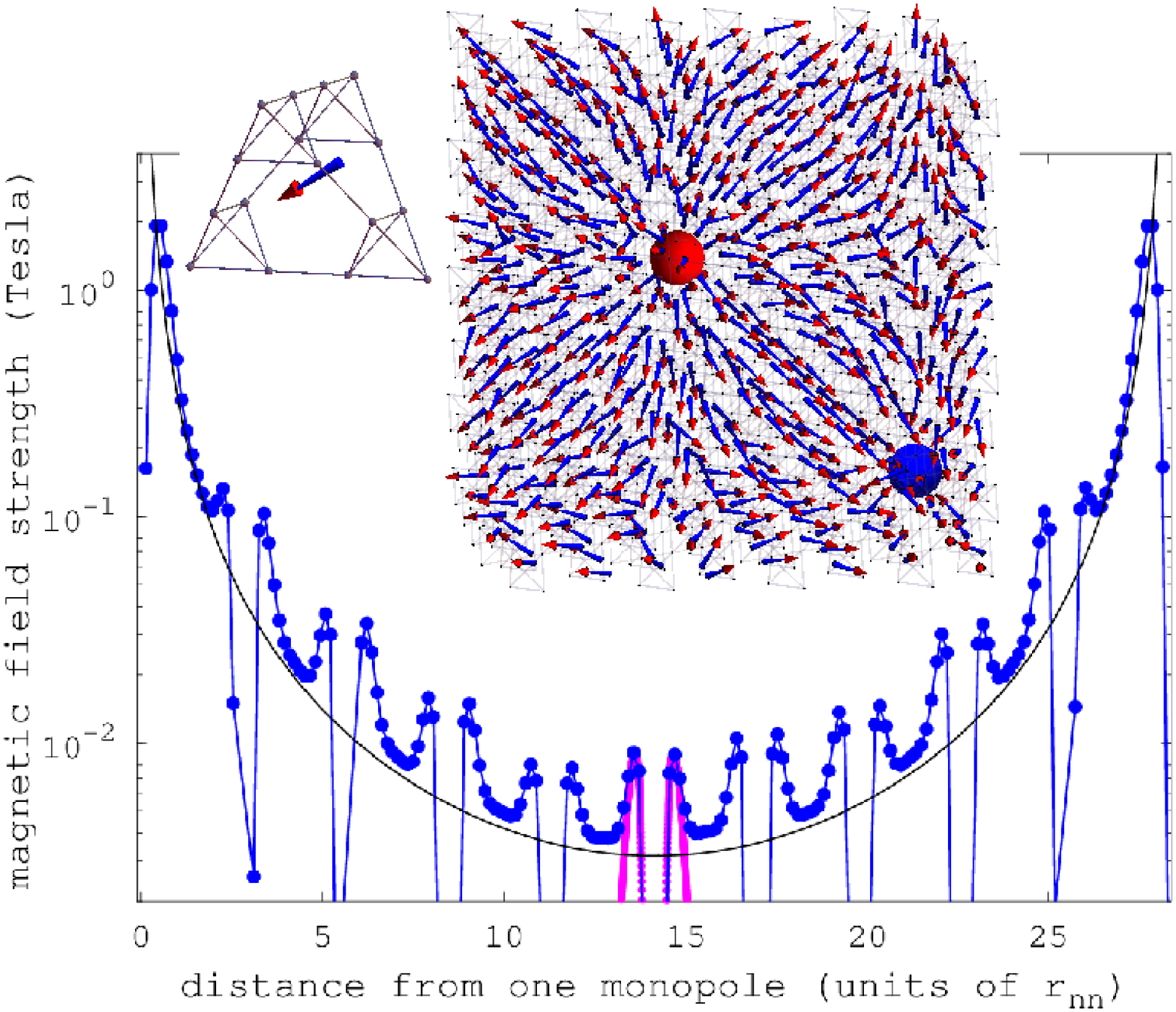}
\caption{
Top:
\label{fig: super tetrahedra}
Illustration of the magnetic field due to the monopoles (red and blue spheres) 
at the centres of the super-tetrahedra of the pyrochlore lattice, visualised
by unit vectors in the local field direction (red-blue arrows), for 1024 spins 
with periodic b.c., averaged over 10,000 independent configurations. 
As shown, each super-tetrahedron is formed by $4$ regular tetrahedra in the 
lattice. The super-tetrahedra centres are (locally) the farthest points from 
any spin on a pyrochlore lattice site. 
Bottom: 
\label{fig: intra mono}
Averaged fields along the $[001]$ line joining two monopoles (connected blue
dots).
The leading behaviour is captured, to within 20\% error, by the Ewald-summed
field from two point magnetic charges at the locations of the monopoles,
with charge from the theoretical prediction~\cite{Castelnovo2008},
$2\mu/a_d \simeq 4.6$~$\mu_B$/{\AA} (black line).
The periodic deviations from the Coulomb form are due to spins which sit very
close to the line -- this contribution is explicitly shown in magenta for the 
spin at the midpoint between the monopoles).
}
\end{figure}

In finer detail, we find that the internal field distribution fits well to a
simple functional form at {\em all} temperatures (Fig.~\ref{fig: histograms}).
Counterintuitively, we find an \emph{enhancement} of the weak field sites as
the temperature is lowered. This is surprising as spin ice is a
ferromagnet -- as defined by the sign of its Weiss temperature -- and one
might expect enhanced internal fields to appear as spins align.
We interpret this unusual behaviour as a result of the interplay of the
nanoscopic lattice structure of spin ice and the slow decay of monopolar
fields.

Overall, our analysis plugs two gaps: firstly, the conceptual one between the 
effective long-wavelength emergent gauge theory~\cite{Castelnovo2008} and the 
nanoscale physics of the lattice; and secondly, the practical one between 
theory and real-space experimental probes. 
%
%

\textit{Distribution of internal field strengths} --- 
In spin ice, the magnetic dipoles reside on the sites of the pyrochlore 
lattice, which consists of corner-sharing tetrahedra. We provide details of 
this structure as Supplemental Information 
but all that is necessary for digesting the following is: 
(i) in any of the exponentially numerous spin ice configurations, two spins 
point into each tetrahedron and two point out; and (ii) an (anti-)monopole 
corresponds to a tetrahedron with three spins pointing in (out). 

In Fig.~\ref{fig: histograms}, we show histograms of the internal field
distribution $P(h)$ collected across the primitive unit cell for three 
different classes of spin configurations 
(see the Methods Section in the Supplemental Information). 
We consider the cases of monopole-free states (red line) and of 
configurations containing two maximally separated monopoles, evaluating the
field in a primitive cell containing a monopole (blue dots) or half way
between the pair (magenta crosses).
This is compared to a random configuration of Ising spins with local
$[111]$ easy axes, corresponding to an infinite temperature state (green).

In all cases, at small fields, $P(h) \propto h^2$, while at large fields,
$P(h) \propto h^{-2}$.
The latter reflects the geometric probability of probing
the $1/r^3$ divergence of $h$ close to a spin.
The former is a non-trivial result which will play an important role in the
interpretation of $\mu$SR experiments further below; 
it implies that a site with a vanishing field is not `special' in the sense
that even an entirely flat probability distribution for each of the three
components of the field vector would yield this functional form for 
$P(h)$.

The presence of a monopole is only weakly visible far away from it but
nearby its effect is felt strongly -- statistical weight is shifted from low
to higher fields, whereas the overall shape of the distribution does not vary
appreciably. This is highly counterintuitive, if one considers that spin ice
2in-2out tetrahedra are \emph{ferromagnetically ordered}: all the spins point
in the same direction, to the extent allowed by the local easy axes.
For a ferromagnet, one would naively expect that the internal fields are 
larger in the 2in-2out arrangement than they are in presence of a monopole 
or otherwise disordered spins.

There are two reasons why this happens nonetheless. Firstly,
the characteristic dipolar correlations between 2in-2out tetrahedra
lead to an unusually large cancellation between fields from different
tetrahedra. The spins form 'flux loops' where the sum of their dipole moments
in fact vanishes. These `flux loops' get broken down as monopoles appear,
whose field decreases with distance {\em more slowly} than that of any dipole.
This effect is captured by the {dumbbell} model~\cite{Castelnovo2008}, which
accounts well for the long-wavelength aspect of the field distribution. 
[A more detailed explanation can be found in a dedicated section in the 
Supplemental Information for this paper.]

However, to reveal the second reason, such a picture needs to be supplemented
to account for the detailed structure of the field distribution on the lattice
scale. This exhibits considerable local
structure, as shown in Fig.~\ref{fig: cross section}:
near the spins, and at the centres of tetrahedra, there are large fields in
excess of $4$~Tesla. By contrast, in the voids between the spins, the fields
average much lower.
Most saliently, at the centres of supertetrahedra (see
Fig.~\ref{fig: super tetrahedra}, inset), the probability of finding a
low-field site is greatly enhanced (black dots,
Fig.~\ref{fig: histograms}).
Indeed, the oscillations in this latter curve provide a crucial
pointer: at these locations, aided by symmetry, the fields of
nearby spins can cancel locally, leaving a lower characteristic field scale,
and hence enhanced low-field probability.

This shows up in  the field distribution averaged across a unit cell
near a monopole (Fig.~\ref{fig: histograms}, magenta line), which follows the
form
$
P(h)
\sim
h^2 / \left(\vphantom{\sum} h^2 + H^2_0 \right)^2
$
derived for the distribution of fields due to randomly located and oriented
spins~\cite{random_fields}. Crucially, the value of $H_0$ is {\em reduced}
compared to that of a defect-free configuration (red line).
This picture is backed up by the good fit of 
the above equation 
to a high-temperature state corresponding to a collection of randomly oriented
$[111]$-easy-axis dipoles (green line), and hence a high density of randomly 
distributed monopoles. 

Finally, Fig.~\ref{fig: cross section} directly demonstrates that, along
with the breaking of ice rules, the spontaneous spatial organisation of
fields strengths into high- and low-field locations within the unit cell is
suppressed.
%
%

\textit{Average field due to magnetic monopoles} --- 
Having analysed the spatial distribution of fields inside the unit cell, we
next turn to visualising the field set up by a monopole. Recall that
magnetic monopoles experience a relative force of Coulombic nature.
We ask: can one also measure the corresponding magnetic field
$(\mu_0/4\pi) (q/r^2)$? This is difficult for two reasons.
Firstly, the internal field away from the lattice sites varies tremendously
between configurations.
Secondly, the `Dirac string'~\cite{Castelnovo2008} emanating
from the monopole carries a magnetisation, $\vec{M}$, which cancels off the
field, $\vec{H}$, to give a net
$
\vec{\nabla}\cdot\vec{B}
=
\mu_0 (
\vec{\nabla}\cdot\vec{H}
+
\vec{\nabla}\cdot\vec{M}
) = 0
$.
These two issues can be taken care of by (i) averaging over many
configurations (or, in experiment, over time whilst keeping the observed
monopole position fixed) and (ii) measuring the field at points as far away
as possible from any lattice sites. (This also minimises the strong
near-field of the spins.)

In Fig.~\ref{fig: super tetrahedra}, we display the direction of the average
fields at the centres of the supertetrahedra set up by two stationary
monopoles, which visually reproduce the expected hedgehog-like monopolar field 
pattern. We have verified that by subtracting the analytical Ewald-summed 
field for two point monopoles with charge given in 
Ref.~\onlinecite{Castelnovo2008}, the residual fields appear randomly oriented 
with average strength tenfold suppressed. 

To be more quantitative, in Fig.~\ref{fig: intra mono} we
show the average field (over $10^6$ configurations) evaluated along a line
joining the two monopoles halfway across a system of $128,000$ spins
($L= 20$), along the $[001]$ direction.
The Coulomb field
predictions are borne out, but masked in part by the line passing very
close to spins, which contribute a strong and periodic deviation (magenta line)
from the theoretical curve at positions $2 \sqrt{2} \, n$, $n = 0, \dots, 10$ 
(in units of $\rnn$). 

It is tempting to speculate that, if a quantum spin ice material were to be
discovered where monopole motion can be made slower than quantum dynamics of
spin-rearrangements preserving the ice rules, the internal fields would be due
almost entirely to the emergent magnetic monopoles!
%
%

%
%

\textit{Experiment I: Zero-field NMR at the O(1) sites} --- 
Given the ferromagnetic interactions, the centres of the rare earth tetrahedra
experience a large internal field of several Tesla (from numerical
simulations using Ewald-summed interactions and fields -- 
see the Methods Section in the Supplemental Information 
for further details), which as we show next can
be used to distinguish between tetrahedra which host monopoles, and
those which do not.

The centres of the rare earth tetrahedra in the pyrochlore lattice are
occupied by oxygen ions (customarily referred to as O(1) oxygens).
The $^{17}O$ isotopes are NMR active and can be used to detect the internal
fields in a zero field NMR measurement. Indeed, two of the present co-authors
have carried out the first NMR experiments on the `monopole-free' line
{as shown in Fig.~\ref{fig: O(1) fields}.} Also shown is a
comparison to Monte Carlo simulations in thermal equilibrium.
The agreement of the lineshape is remarkably good.
{The experimental spectrum is actually peaked at $3.4$~Tesla,
i.e., about 25\% below the calculated peak.
This shift is most likely due to spatial distribution of Dy-4f electrons
causing deviations from a point dipole approximation (multipolar effects)
and/or effects of Dy-O chemical bonding.}

A promising aspect of these NMR measurements lies in the possibility of
direct detection of monopoles. Indeed, when a tetrahedron hosts a monopole,
the field at its O(1) site drops by approximately $13$\%.
This effect is $3$ times larger than the linewidth due to variations in the
field at the O(1) site because of neighbouring monopoles or more distant
spins. Therefore, the relative intensity of the NMR signal at this pair of
field values provides a quantitative
measure of the density of monopoles. In practice, small densities may be
hard to detect above the background.
{Also, monopoles must not move over the time scale scale of NMR
spin-echo experiments (a few tens of microseconds) to be detected as a
distinct resonance line. Given the insights from modelling AC susceptibility
results, which suggest a hopping rate in the range of
milliseconds~\cite{Snyder2004,Jaubert2009}, this condition seems
comfortably achievable.}
%
%

\textit{Experiment II: $\mu$SR} --- 
The other major probe of the local magnetic fields is $\mu$SR. Indeed,
Ref.~\onlinecite{Bramwell2009} has reported using $\mu$SR experiments to
measure the monopole charge via an ingenious analogy to the Wien effect
familiar from electrolytes. This is currently a contested
experiment \cite{Dunsiger2011} and we would like to make a few observations
on it from the perspective of the current paper. 

First, we note that one important feature of the data presented in
Ref.~\onlinecite{Bramwell2009} is that their signal was extracted from
sites where the muons experience very low fields, of order a few
milliTesla. This follows from the observation that doubling
a transverse field of strength $1$~mTesla doubles the $\mu$SR precession
frequency. From our work it is clear that there is {\em a very low density}
of such sites (see Fig.~\ref{fig: histograms}), located preferentially near
the centres of supertetrahedra (Fig.~\ref{fig: cross section}). It thus
seems very unlikely that the substantial muon signal emanates from sites in
the pristine bulk.

Second, the analysis of Ref.~\onlinecite{Bramwell2009} invokes two distinct
phenomena: first, the applied magnetic field induces an increase in the
monopole density; and second, this increase leads to an enhanced 
depolarization rate.
The first feature is indeed an expected phenomenon for the steady state of
electrolytes in a field, but for spin ice, it can only occur as a transient:
since monopole motion magnetises the sample, there can be no steady state
with a nonzero monopole current. Moreover, in thermal equilibrium, the
density of monopoles is believed to decrease in an applied field. About this
our present work has nothing to say.

However regarding the second step, our results indicate that a field-induced
regime with heightened monopole density would be accompanied by a depletion 
of low-field sites. At least this aspect is qualitatively in
keeping with the finding in Ref.~\onlinecite{Bramwell2009} that the rate of
decay of the $\mu$SR asymmetry gets larger in the Wien setting as the
applied field is increased.

Third, combining the above observations leads us to consider the 
interesting possibility that the signal arises from the action of an enhanced 
monopole density on muons implanted {\it outside} the sample. 
The idea here is that outside the sample the monopole fields would dominate
over the much smaller fields present in their absence---given that the field 
of a monopole decays less slowly than that of an isolated spin. 
[Interestingly, it was very recently suggested in 
Ref.~\onlinecite{Dunsiger2011} 
that the signal from muons inside the sample is lost altogether because of 
large (compared to $1$~mTesla) and fast magnetic fluctuations and that the 
only measurable signal due to spin ice comes from muons implanted outside, 
sensitive to stray fields, which are analysed in 
Ref.~\onlinecite{Blundell2011}.] 

A very rough estimate based on a monopole liquid subject to Debye
screening~\cite{Castelnovo2011} suggests that in the temperature range
$T \sim 0.2-0.5$~K, the magnetic field set up by a monopole measured a 
distance roughly $100\pm50$~{\AA} from the sample surface both lies in the 
range relevant for $\mu$SR (between $0.1$ and $1$~mTesla) and dominates that 
set up by an individual spin in the sample. 
Note that -- unlike stray fields set up by the magnetisation induced by a 
uniform external field -- the monopole density grows with temperature, thus 
providing a qualitative discriminant between the two. 
A combined study of temperature-dependent (uniform) 
susceptibility and $\mu$SR experiments therefore looks like the most 
prominent direction to make progress on this issue. 

All such considerations point to the need for more detailed studies of what
happens near the surface of a sample~\cite{Ryzhkin2011} -- e.g.,
what happens to the surface monopole density (as a function of time and
field)? Moreover, little is known at present concerning the surface of
spin ice samples, e.g., what the nature of the local crystal fields is or how
the magnetic lattice terminates. It is also worth bearing in mind that 
equilibration in spin ice may be incomplete at low temperatures, as the 
equilibrium monopole density vanishes exponentially -- at $70$~mK, the lowest 
temperature accessed so far in $\mu$SR measurements, their number density is 
estimated to be such that there is only one pair for a {\em macroscopic} 
volume of a quarter of a cubic metre! 

Finally, we would be remiss if we did not note that our considerations here
have entirely to do with spatially fluctuating static fields and their 
dephasing of muon precession, whereas Ref.~\onlinecite{Dunsiger2011} proposes 
that large dynamical fluctuations are present. The tension between their 
proposal and the magnetic susceptibility data that are consistent with much 
slower dynamics for the spins will need to be resolved before a consistent 
account of spin dynamics in spin ice can be given. 
%
%

\textit{Conclusions} --- 
We have provided a way of visualising the magnetic field of a magnetic 
monopole inside spin ice, yielding the first nanoscopic real space picture of 
this fractionalised  excitation. Our work attests to the reality of the 
monopolar magnetic field not only at long wavelengths but also on the lattice 
scale.

As experimental proofs of the existence of monopoles move in the direction 
from thermodynamics towards increasingly microscopic `single-monopole' 
detection, we hope that this work will lay the theoretical groundwork for 
future searches, such as the ones using NMR or local field probes outside 
the sample which we have outlined above.
%
%

\textit{Acknowledgments} --- 
We thank Steves Blundell and Bramwell, Sarah Dunsiger, Sean Giblin,
Chris Henley, Jorge Quintanilla, and Tomo Uemura for several useful
discussions.

G.S. and C.C. are grateful to ISIS at the Rutherford Appleton Laboratories
for hospitality and financial support.
This work was supported in part by EPSRC Postdoctoral Research Fellowship
EP/G049394/1 (C.C.).
We mutually acknowledge hospitality and travel support
for reciprocal visits.
%
%

%
%

\section{\label{sec: suppl info}
Supplemental information
        }
%
%

\subsection{\label{sec: spin ice}
Spin ice basics
           }
Here, we briefly collate some basic facts about spin ice -- the reader 
interested in a more detailed and complete overview is referred to 
Ref.~\onlinecite{Bramwell2001-1} and the forthcoming 
Ref.~\onlinecite{Castelnovo2012-1}. 

\begin{figure}[ht]
\includegraphics[width=0.49\columnwidth]
                {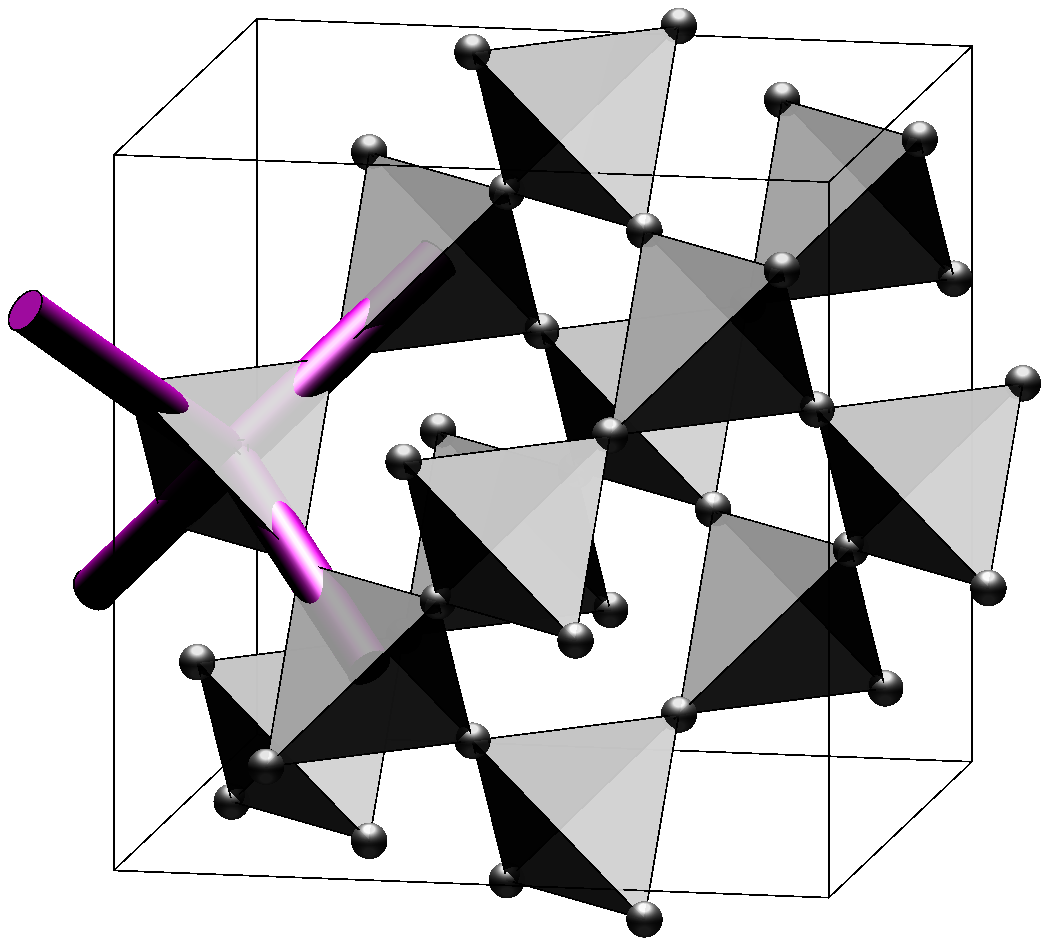}
\includegraphics[width=0.49\columnwidth]
                {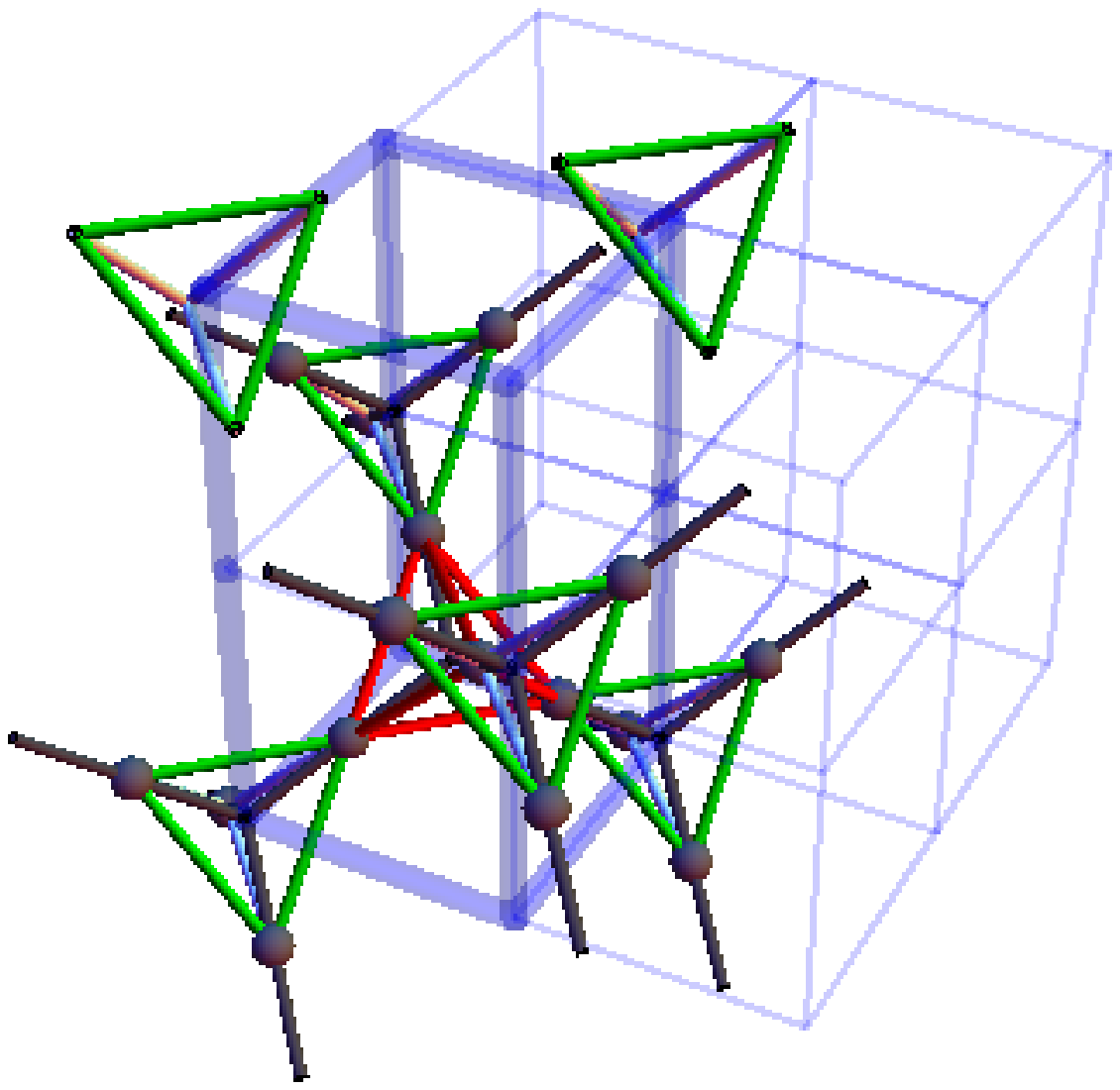}
\caption{
\label{fig: primitive cell}
Left panel: Illustration of the pyrochlore lattice. 
Right panel: 
Conventional cubic unit cell in spin ice, containing $16$ spins (dark gray
spheres). The side of the cell equals $2 \sqrt{2} \rnn$, where $\rnn$ is
the pyrochlore lattice nearests-neighbour distance.  A primitive unit cell 
(highlighted by thick blue
lines) is 4 times smaller than the cubic unit cell and contains only $4$
spins. The black lines are the bonds of the diamond lattice formed by the
centres of the tetrahedra and they identify the local $[111]$ axes of the
rare earth magnetic moments. The distance between the centres of two
neighbouring tetrahedra is given by the diamond lattice constant,
$a_d = \sqrt{3/2}\,\rnn$.
}
\end{figure}

The canonical spin ice materials are {\DTO} and {\HTO}. The magnetic ions 
Dy$^{3+}$ and Ho$^{3+}$ live on a lattice of corner sharing tetrahedra 
(pyrochlore lattice) and they have large effective spins: $J=15/2$ and $J=8$, 
respectively. A strong local easy axis anisotropy forces the spins to lie 
along the lines connecting the centres of adjacent tetrahedra (namely, the 
local $[111]$ axis, see Fig.~\ref{fig: primitive cell}). 
The combination of large electronic spins and strong easy axes leads to 
large magnetic moments of the order of $10$ Bohr magnetons per spin. 
The long range dipolar interactions are dominant. 
To a good approximation, the full Hamiltonian of these systems can be 
written as 
\bea
H
&=&
J_{\rm ex} 
\sum_{\langle i , j \rangle}
  {\bf S}_i \cdot {\bf S}_j
\nonumber \\ 
&+&
\frac{\mu_0 \mu^2}{4 \pi} \sum_{i < j}
\left[
  \frac{{\bf S}_i \cdot {\bf S}_j}{ r_{ij}^3}
  -
  \frac{3({\bf S}_i \cdot {\bf r}_{ij})({\bf S}_j \cdot {\bf r}_{ij})}
       {r_{ij}^5}
\right]
,
\label{eq:dipH}
\eea
where the summation is over the indices $i,j$ of the spins on the lattice, 
$J_{\rm ex} > 0$ is the exchange coupling constant, $\mu_0$ is the vacuum 
permeability, and $\mu \simeq 10 \mu_B$ is the magnetic dipole moment of 
a spin. 

The interplay between the strong local easy axes and the 
dipolar interactions favours nearest-neighbour `pseudospin 
antiferromagnetism': a spin pointing into a given tetrahedron wishes its 
neighbors on the same tetrahedron to point out and vice versa. The
resulting ground-state configuration are described by the ice rules,
which stipulate that two spins point in and two point 
out of each tetrahedron. Exponentially many configurations satisfy the ice
rules, leading to an {\em extensive} low-temperature entropy of 
$S_p\approx\frac12\log\frac32$, as verified experimentally 
by Ramirez \etal~\cite{Ramirez1999}. 
Hence, despite a ferromagnetic 
Curie-Weiss temperature $\approx 2K$, the 
spin ice compounds fail to develop long-range spin order down to a 
temperature four times lower. 

This peculiar degeneracy has two remarkable consequences. Most basically, 
the spin correlators are characterised by an emergent gauge symmetry,
which results from an effective conservation law for the magnetisation density 
which is implied by the ice rules--the effective theory of spin ice is an 
emergent magnetostatics. 
In such a  theory, the elementary excitations in spin ice do not take the 
conventional form of extended domain walls. Rather, flipping a spin in a 
2in-2out state generates two defective tetrahedra which have 3in-1out and 
3out-1in spins, respectively. 
These defects are then free to move through the system by means of spin 
flip processes that do not generate further defects: the initial dipolar 
perturbation (a spin flip) fractionalises into two point like excitations 
(the two defective tetrahedra). 
Indeed, one can formally show that these excitations are truly 
deconfined~\cite{Castelnovo2008-1}, being subjected only to a Coulombic ($1/r$) 
interaction potential. 

These defects can be shown to carry 
a net magnetic charge, so that spin ice is the first experimentally 
accessible setting where free point-like magnetic charges can be observed 
and manipulated. 
%

\subsection{\label{sec: flux loops}
Flux loops and temperature dependence of the internal fields
           }
In order to understand this counterintuitive temperature dependence
of the distribution of low field sites in spin ice materials, it is
useful to note that this system is very unusual in that it is a magnet
in which {\em ferro}magnetism appears in a frustrated way. 

{\em (1) Ferromagnetism:} at a single tetrahedron level, the dipolar
interactions are minimised by a ferromagnetic arrangement of the
spins. The well-known 2in-2out ice rules maximise indeed the magnetic
moment of a tetrahedron. {\em (2) Frustration:} the
degeneracy resulting from the frustration-induced underdetermination
leads to longer range correlations, where the local moments do not
align, like in a usual ferromagnet, and hence do not produce a
macroscopic moment. Rather, the (effective) emergent gauge structure
is one which suppresses the coarse-grained moments locally.~\cite{Huse2003} 
Pictorially and intuitively (if not rigorously), this can be understood 
by the dipoles aligning head-to-tail to form strings which have a preference 
for closing back on themselves. 
Such strings exhibit zero net mono- and dipole moments in a continuum 
treatment, which in turn leads to an effective suppression of the fields 
they produce at large distances. 

When monopoles are present, the formation of the closed loops breaks
down as monopoles introduce endpoints of such loops. [An open loop
terminated by monopoles is commonly referred to as a 'Dirac string' in
the literature.]  Therefore, the rapid decay of the loop field due to the
vanishing of the monopole and dipole moments is replaced by the slow decay 
of the Coulomb field. 

The fields due to a distribution of monopoles thus combine to give an 
enhanced typical field strength, and hence a concomitant reduction in the 
density of low-field sites. This explains the observed effect. 

Indeed, this behaviour is immediately apparent in the dumbbell model
introduced in Ref.~\onlinecite{Castelnovo2008-1}. 
In that model, each spin is replaced by a
dumbbell of charges located at the opposite ends of the lattice vector
separating the centres of two adjacent tetrahedra. When the ice
rules are satisfied throughout the system, two positive and two
negative charges overlap at the centre of each tetrahedron. This leads
to a charge neutral system that gives rise to vanishing internal
fields. Adding monopoles replaces the effectively charge-neutral
field-free system with one in which charges are distributed
throughout, leading to a complex superposition of `stray fields' of
all the monopoles at any given site of the lattice. 

These arguments neglect features of the short-distance behaviour,
which as we show in this work are themselves non-trivial. In
particular, lattice-scale variations of the field are substantial,
ensuring the (energetically required) decrease of the effective field
strength at the sites of the spins as monopoles are created with
increasing temperature. 

Now, low-field sites occur due to the cancellation of the
strong fields of nearby spins. The probability of such local
cancellation is not much influenced by the presence of a distant
monopole; however, even if such local cancellation takes place, 
it is then spoilt by the above mentioned monopolar `stray fields'. 
Hence, the occurrence of low-field sites is suppressed by the creation 
of monopoles. 
%
%

\subsection{\label{sec: methods}
Methods
           }
Statistically independent spin configurations were obtained by means of loop
updates that do not introduce violations to the 2in-2out ice rules and
do not alter the location or charge of existing monopoles in the system.
The internal fields for each configuration were obtained using the Ewald
summation technique~\cite{Ewald_refs}, 
with the {\DTO} parameters given in Ref.~\onlinecite{denHertog2000}.

The average field strengths in 
Fig.~1 in the manuscript 
and the
histograms in 
Fig.~3 in the manuscript 
were obtained from a uniform cubic
grid of $16,000$ points (corresponding to a point spacing of $< 0.25$~{\AA})
across a primitive unit cell of the system (Fig.~\ref{fig: primitive cell}),
averaged over $\sim 10^4$ realisations.

The Monte Carlo simulations that produced the distribution in
Fig.~2 in the manuscript 
used single spin flip updates and Ewald-summed
dipolar interactions between the spins, for a system of $3,456$ spins
($L = 6$).

The O-NMR experiments requires $^{17}$O, which is the only oxygen isotope with
a non-zero nuclear magnetic moment. We have enriched $^{17}$O content in a crystal of
Dy$_2$Ti$_2$O$_7$ grown by a floating-zone image furnace from the natural
abundance of 0.01\% to approximately 50\% by annealing the crystal under 65\% 
$^{17}$O$_2$ gas at 1000$^\circ$C for several days. Only the  O(1) sites at
the 8a positions located at the centers of the Dy-tetrahedra experience large
internal fields $\gtrsim 3$~Tesla, while the internal fields at the O(2) sites
at the 48f positions are much smaller and do not contribute to the observed
signal.
Moreover the electric field gradient is zero at the O(1) sites. Therefore,
the NMR spectrum at the O(1) site represents the distribution of the local
fields (directly obtained by dividing the frequency by
$5.77185$~Tesla/MHz, which is the gyromagnetic ratio of $^{17}$O nuclei).
To construct a broad-band spectrum, each Fourier spectrum of a pulsed
NMR echo is summed while the frequency is being swept (Fourier-step-summing
technique). Low-temperature experiments have been carried out by putting the
sample into a one-shot liquid $^{3}$He fridge or into a dilution
fridge using a He mixture.
%
%

\subsection{\label{sec: further experiments}
Further experiments
           }
%
%

\subsubsection{\label{sec: experiments2}
Surface experiments
              }
Ideally, we would like to measure the monopole field
(Fig.~4 in the manuscript) 
by placing probes
at the centres of supertetrahedra. Interestingly, these coincide with the
centres of tetrahedra of titanium ions. Placing an NMR-active ion there
would in principle allow access to the Coulombic form of the internal fields.

However, even if this remains a thought experiment for the time being,
it is clear that a promising strategy is to suppress short-distance
fluctuations by moving away from spin locations even on the scale of a
lattice constant 
(Fig.~4 in the manuscript).
The obvious way to do this is to probe the fields close to the surface of a
sample, as discussed in the previous section. The spins and strings being
confined to the sample, one can thus hope
to see the farther-ranging monopole magnetic field undisturbed.
Indeed, it is such a set-up that was proposed by Zhang~\etal~\cite{Zhang2009}
to measure the image magnetic monopole induced by an electric charge near
the surface of a strong topological insulator.
%
%

\subsubsection{\label{sec: avalanches}
Magnetic avalanches 
              }
Our work also sheds  light on a range of non-equilibrium phenomena
recently measured in spin ice~\cite{Slobinsky2010}, 
and their prominent
analogues in artificial spin ice~\cite{Mellado2010,Mengotti2011-1}. 
While the field strength throughout the unit cell fluctuates wildly,
the field projected onto the spin direction on a lattice site
is known to be near constant for an `ideal' interaction 
[S.~V.~Isakov, R.~Moessner, and S.~L.~Sondhi~\cite{Isakov2005}; 
for the conventional dipolar Hamiltonian, we find a broadening to
$0.80 \pm 0.03$~Tesla. This field sets the energy scale for creating
bound monopole pairs at low temperature. This quantity -- as well as the
Coulomb interaction between two monopoles as they separate -- is important
for triggering the avalanches which have received much attention of 
late~\cite{Slobinsky2010,Mellado2010,Mengotti2011-1}. 
%
%

\end{document}